\documentclass[12pt]{article}
\usepackage{amsmath}
\usepackage{amssymb}
\input amssym.def
\input amssym
\newcommand{\be}{\begin{equation}}
\newcommand{\ee}{\end{equation}}

\title{Classification of electromagnetic fields in general
relativity and its physical applications}
\author{Nikolai V. Mitskievich\thanks{Physics Department, CUCEI,
University of Guadalajara, Guadalajara, Jal., Mexico.}
\thanks{Postal address: Apartado Postal 1-2011, C.P. 44100,
Guadalajara, Jalisco, M\'exico. E-mail:
mitskievich03@yahoo.com.mx}}
\date{~}
\begin{document}

\maketitle
\begin{abstract}
The simplest electromagnetic fields' (general- as well as
special-re\-lativistic) classification is formulated which is
based on physically motivated ideas. According to this
classification these fields can belong to three types (electric,
magnetic and null), each of them being split in pure and impure
subtypes. Only pure null type field propagates with the
fundamental velocity $c$, all other fields have the propagation
velocity less than that of light. The reference-frame-based
methods of elimination of alternative three-fields ({\it e.g.},
magnetic in the electric type case) are given for pure subtypes;
for pure null type the generalized Doppler effect takes place
instead. All three types of impure fields are shown to be {\bf
E}-{\bf B}-parallelizable. Thus such an elimination in pure
non-null and parallelization in all impure cases mean
transformation to the reference frame co-moving with the
electromagnetic field in which the Poynting vector vanishes. The
methods we propose modernizing the Rainich--Misner--Wheeler
approach, also permit to construct new exact Einstein--Maxwell
solutions from already known seed solutions. As examples, the
Kerr--Newman and Li\'enard--Wiechert solutions are considered,
three ``new'' types of rotating charged black holes (with the same
Kerr-Newman geometry) are presented, and new physical effects are
evaluated.\\
PACS 2008 Numbers: 04.20-{\bf q}, 04.20.Ex, 04.40.Nr, 04.70.Bw
\end{abstract}

\renewcommand{\theequation}{\arabic{section}.\arabic{equation}}

\section{Introduction} \label{s1}

The main idea of this paper is to present an elementary practical
classification of electromagnetic fields (see also
\cite{Mitsk06a}) equally applicable in both relativities and
having deep physical roots. In fact, there already exist
classifications of stress-energy tensors (essentially, those of
Segr\`e and Pleba\'nski, \cite{Pleb,Segre,Exact2}) and,
specifically, of electromagnetic fields \cite{SyngeSR}. However,
one does not encounter there a direct relation to such field
properties as the Poynting vector and the velocities of
propagation of concrete configurations of these fields (in
particular, with respect to the reference frames co-moving with
the corresponding fields when these velocities are $<c=1$ in the
natural units used here; for a detailed discussion of the
Li\'enard--Wiechert example see our recent article \cite{Mitsk06a}
and a preprint with complete simple deduction of that solution
\cite{Mitsk05}). Now, our aim is also to give here natural and
simple prescriptions for calculation of such physical
characteristics of every concrete electromagnetic field ({\it cf.}
\cite{MisWh,Rain1,Whee}).

We consider electromagnetic fields in a vacuum (but electric
currents may be present, not the media with dielectric and
magnetic properties), this theory being treated here on the
classical (not quantum theoretical) basis. The state of motion (in
general, inhomogeneous) of the electromagnetic field, which we
call its propagation, is that of a concrete (preferably, exact)
solution of the system of dynamical equations (of Maxwell in
special theory and Einstein--Maxwell in general theory of
relativity), but not the propagation of perturbations on the
background of these solutions, and not the propagation of
discontinuities, which belong to other problems of the field
theory not considered in this paper (their treatment is already
well developed and does not need immediate revision). Any motion
is, naturally, relative (that occurring with the velocity of light
is also relative, at least in the sense of its direction: the
light aberration effect), while motions with under-luminal
velocities always permit to introduce co-moving reference frames
with respect to which the fields are ``at rest'' everywhere in the
four-dimensional region of the frame determination, which includes
the requirement that redistributions of the field
(``deformations'') should be ``caused'' by deformation,
acceleration and rotation of the respective co-moving reference
frame, and {\it vice versa}. In this connection, we remind that
the physical reference frame is an idealized image of a changing
with time (let us not to put this concept here more precisely)
distribution and motion of observers together with their
observational and measuring devices, idealized primarily in the
sense to be test objects, {\it i.e.} they should not practically
perturb the characteristics of the objects (including fields and
spacetime) in the classical (non-quantum) theory. We do not touch
here upon the problem of quantum theoretical description of
reference frames: this question still seems not to be adequately
considered in physics, although the first step in this direction
was already made by Bohr and Rosenfeld \cite{BohrRos}.

\subsection{A preview of the paper's structure} \label{s1.1}

The reader will notice that in this paper the original conclusions
are enlarged with a modernized review of already known facts to
make the exposition more self-sufficient. In the next section the
notations and definitions used in this paper are given. In section
\ref{s3} we present a condensed information on description of
reference frames and its application to electromagnetic fields.
The short section \ref{s4} is dedicated to the classification of
these fields in terms of their two invariants, $I_1$ and $I_2$. In
section \ref{s5} we introduce the concept of propagation velocity
of electromagnetic fields in a vacuum, and it is shown that
absolute value the three-velocity of all pure null fields is equal
to unity. The pure electromagnetic fields (when $I_2=0$) are
considered in section \ref{s6} yielding for the pure electric and
magnetic types a simple elimination of magnetic or electric field,
respectively, by the corresponding choices of reference frame
(subsection \ref{s6.1}), and a specific r\^ole of the Doppler
effect (with its inevitable generalization) for the pure null type
in subsection \ref{s6.2}. An approach to constructing exact
Einstein--Maxwell solutions in the same 4-geometry as that of any
exact seed Einstein--Maxwell solution one arbitrarily would choose
(with the exception of pure null fields), is developed in sections
\ref{s7} and \ref{s8}. Section \ref{s9} is dedicated to the
treatment of impure subtype ($I_2\neq 0$) of all three types of
fields leading to parallelization of electric and magnetic vectors
in the adequate (canonical) reference frames. In sections
\ref{s10} and \ref{s11} we consider application of the methods
developed in the preceding sections to some exact solutions of the
Einstein--Maxwell equations in general relativity and Maxwell's
equations in special relativity: in subsection \ref{s10.1} of the
well-known Kerr--Newman (KN) solution (involving, as we show, in
different regions different types of electromagnetic field whose
electric and magnetic vectors are always collinear and
radially-directed), and in subsection \ref{s11.1}, the
Li\'enard--Wiechert solution (we show that it belongs to the pure
electric type and there always exists a global co-moving with this
electromagnetic field non-degenerate reference frame, so that the
velocity of propagation of this field in a vacuum is everywhere
less than that of light, with the exception of the future null
infinity). In subsection \ref{s10.2} ``new'' exact solutions with
pure electric and pure magnetic fields in the standard Kerr-Newman
black hole geometry are presented, together with a similar black
hole with impure-null-type electromagnetic field. In subsection
\ref{s11.2}, it is shown that a superposition of plane harmonic
electromagnetic wave and homogeneous magnetic field has strictly
sub-luminal velocity of its propagation in a vacuum. In the final
section \ref{s12} the obtained results are summed up and
concluding remarks are given.

\section{Mathematical preliminaries} \label{s2}
\setcounter{equation}{0}

Everything will be considered in four spacetime dimensions. We use
the spacetime signature $+,-,-,-$, Greek indices being
four-dimen\-sional (running from 0 to 3), and Latin ones,
three-dimensional, with the Einstein convention of summation over
dummy indices. However, in the reference frame formalism, all
indices usually are Greek, and the splitting into physical
spacelike and timelike objects only means that the former ones are
in all free indices orthogonal to the timelike monad vector
(projected onto the physical three-space of the reference frame),
while the physical timelike parts represent contractions with the
monad in the indices which hence become absent (a change of the
root-letter notation is then advisable). The indices put into
individual parentheses belong to tetrad components.

For the sake of convenience and writing and reading economy, the
Cartan exterior forms formalism is frequently used. In it, the
coordinated basis is the set of four covectors (1-forms) $dx^0,
\dots,dx^3$, and the orthonormal tetrad basis similarly is
$\theta^{(0)}, \dots,\theta^{(3)}$. Every such basis 1-form, ({\it
e.g.}, $dx^2$, $\theta^{(3)}$), itself represents an individual
four-dimensional covector. The exterior (wedge) product simply is
a skew-symmetrized tensorial product (antisymmetrization is also
denoted by Bach's square brackets which embrace the indices, while
factor $\frac{1}{(\textnormal{their number})!}$ is supposed to be
included in this definition). It is clear that the rank of a form
can be from 0 (a scalar) to 4, inclusively; all forms of higher
ranks vanish identically (in $D=4$). The scalar product of
four-vectors or covectors is denoted by a central dot, if these
vectors are written without indices ($A\cdot B$), but {\it with}
such indices this has a wider meaning, for example, $dx^\mu\cdot
dx^\nu=g^{\mu\nu}$: literally, this means that the scalar product
of two coordinated-basis covectors equals a contravariant
component of the metric tensor with the same indices as those of
these factors.

The dual conjugation in the sense of components (their indices) is
denoted by an asterisk over the corresponding subindices, or under
upper indices; the Hodge star stands for dual conjugation of a
form written more abstractly, and is denoted by an asterisk before
the form; it is convenient to have in mind that, after all, it
applies to the form's basis, though this is equivalent to a
similar dual conjugation of the form's components ({\it not both
at once!}). An application of a pair of Hodge stars does not
change an odd-rank form and results in the change of the sign of
an even-rank form (for example, the electromagnetic field 2-form
$F$). By this {\it definition}, \be \label{dual}
\ast(dx^{\alpha_1} \wedge\dots\wedge
dx^{\alpha_k}):=\frac{1}{(4-p)!}{E^{\alpha_1
\dots\alpha_k}}_{\beta_1\dots\beta_l}(dx^{\beta_1}\wedge\dots
\wedge dx^{\beta_l}) \ee where \be \label{E} E_{\kappa\lambda\mu
\nu}:=\sqrt{-g}\epsilon_{\kappa\lambda\mu \nu}, ~ ~ E^{\kappa
\lambda \mu\nu}:=-\frac{1}{\sqrt{-g}}\epsilon_{\kappa\lambda\mu
\nu} \ee are covariant and contravariant components of the axial
Levi-Civit\`a tensor, and the usual Levi-Civit\`a symbol is
defined as \be \label{LC} \epsilon_{\kappa\lambda\mu \nu}=
\epsilon_{[\kappa\lambda\mu \nu]}, ~ ~ \epsilon_{0123}=+1 \ee
(always with the {\it sub}\,indices: this is a {\it symbol},
though simultaneously representing components of a contravariant
axial tensor density of the weight $-1$ {\it and} a covariant
axial tensor density of the weight $+1$). See some details in the
beginning of the introductive chapter in \cite{Mitsk06}. Finally,
coming back to a formula in the end of the second paragraph of
this section, we have $\ast(dx^\mu\wedge\ast dx^\nu)=-dx^\mu\cdot
dx^\nu=-g^{\mu\nu}$. Of course, the r\^ole of metric properties of
spacetime is somewhat hidden in the Hodge notations, as one can
see from the formulae (\ref{dual}) and (\ref{E}).

\section{Algebra of reference frames; applications to the
electromagnetic field} \label{s3} \setcounter{equation}{0}

The central point of our paper is the use of algebraic
considerations, other applications being here only of auxiliary
significance, for example, the exterior differentiation operator
$d=\theta^{(\alpha)} \nabla_{X_{(\alpha)}}\wedge\equiv
dx^\alpha\nabla_{\partial_\alpha}\wedge$.

In the physical sense, a concrete reference frame (see
\cite{Mitsk06}) has only to do with a state of motion (a timelike
world lines' congruence, or, equivalently, its unitary tangent
vector field, the monad $\tau$) of a swarm of test observers
together with their test measuring devices. Moreover, one
additional ingredient, the metric tensor $g$, is needed to
construct the projector $b:=g-\tau\otimes\tau$ which at the same
time serves as the (formally, four-dimensional) metric tensor on
the three-dimensional local subspace orthogonal to the monad field
$\tau$; $b_{\mu\nu}\tau^\nu\equiv 0$, $\det b\equiv 0$. $b$ has
the signature $0,-, -,-$, so that the ``three-dimensional'' scalar
product of two vectors is \be \label{bullet} A\bullet B:=-
b_{\alpha\beta}A^\alpha B^\beta\equiv\ast[(\tau\wedge A)\wedge\ast
(\tau\wedge B)] \ee where these vectors are also automatically
projected onto the local subspace mentioned above. If such vectors
did already belong to the subspace, they usually are boldfaced:
$\mathbf{A}^\mu =b^\mu_\nu A^\nu$. The ``three-dimensional'' axial
vector product of two vectors now reads \be \label{times} A\times
B:=\ast(A\wedge\tau\wedge B). \ee These algebraic operations are
locally equivalent to the usual three-dimensio\-nal scalar and
vector products, so we denote them by essentially the same
symbols. In fact, in the complete reference frame theory we
similarly use the operations of gradient, divergence and curl,
but, being differential operators, they are more profoundly
generalized, explicitly taking into account the characteristics of
inhomogeneities of general reference frames, such as acceleration,
rotation and deformation (expansion and shear) which naturally
cannot be present in the algebraic treatment of geometry. The
uniformity of general and old traditional notations radically
simplifies the physical interpretation of general- and (in
non-inertial frames) special-relativistic expressions as well as
of theoretically predicted effects.

Electromagnetic fields are described with the use of the covector
potential $A=A_\alpha dx^\alpha$ and the 2-form (the field tensor)
\be \label{F2form} F=dA=\frac{1}{2}F_{\alpha\beta}dx^\alpha\wedge
dx^\beta. \ee With respect to a given reference frame $\tau$ (see
\cite{Mitsk06}), the field tensor splits into two four-dimensional
(co)vectors, electric \be \label{elE} \textnormal{{\bf
E}}_\mu=F_{\mu\nu} \tau^\nu ~ ~ \Longleftrightarrow ~ ~
\textnormal{{\bf E}}= \ast(\tau \wedge\ast F) \ee and magnetic \be
\label{magB} \textnormal{{\bf
B}}_\mu=-F\!\!\stackrel{\textnormal{\small$\ast$}
}{\textnormal{\scriptsize$\mu\nu$}}\!\tau^\nu ~ ~
\Longleftrightarrow ~ ~ \textnormal{{\bf B}}=\ast(\tau\wedge F),
\ee both $\perp\tau$, thus \be \label{FBE} F=\mathbf{E}\wedge\tau+
\ast(\mathbf{B}\wedge\tau). \ee It is obvious that {\bf E} is a
polar four-vector and {\bf B}, an axial four-vector, both
restricted to the local physical three-subspace of the
$\tau$-reference frame. (In Cartesian coordinates and with the
corresponding inertial monad, consequently, in the Minkowskian
spacetime, we have the same relations as for usual contravariant
three-vectors: $ \textnormal{{\bf E}}^i=F_{i0}=-F^{i0}, ~ ~
\textnormal{{\bf B}}^i =-\frac{1}{2}
\epsilon_{ijk}F_{jk}=-\frac{1}{2} \epsilon_{ijk}F^{jk}$.) The
splitting (\ref{elE}), (\ref{magB}), (\ref{FBE}) follows from the
observation that the Lorentz force can be expressed as \be
\label{LorForce} (\textnormal{{\bf E}}+\textnormal{{\bf v}}\times
\textnormal{{\bf B}})_\alpha=F_{\mu\nu}\left(\tau^\nu+
\textnormal{{\bf v}}^\nu\right) b^\mu_\alpha. \ee

Here the three-velocity of the charged particle on which acts the
Lorentz force, follows from the general definition \be \label{v}
u= \stackrel{(\tau)}{u}(\tau+\mathbf{v}) ~ \Rightarrow ~
\textnormal{\bf v}=b(\frac{dx}{dt},\cdot) \ee where
$\stackrel{(\tau)}{u}=u\cdot\tau=\frac{dt}{ds}=(1-v^2)^{-1/2}$,
while $dt= \tau_\mu dx^\mu$ ($\tau\cdot dx$, that is non-total
differential of the physical time along an infinitesimal
displacement of the particle in spacetime), and $u$ is its
four-velocity. We have to add here important comments related to
the basic concepts of both relativities, and these comments could
be more transparent just with the three-velocity as an intuitively
clear example. The ``physical'' objects (such as $\mathbf{v}$,
$\mathbf{E}$, {\it etc.}) belong to the section orthogonal to
$\tau$ using which these objects are introduced. Thus, already in
special relativity, the velocities considered in their composition
law, may exist even in three distinct sections of spacetime, while
three frames are participating in the composition, and it is
absolutely obvious that one cannot simply add vectors from two
subspaces obtaining the third one automatically lying in the third
subspace, all of them having necessary properties with respect to
these respective frames. And in the composition law the
three-vectors being added together, frequently are ``collinear''
(an absurd if they belong to non-parallel sections of spacetime).
This is, of course, understandable, since Einstein himself did not
realize the fact of unification of space and time into the
four-dimensional manifold before the famous discovery of Minkowski
in 1908 (and even during several years after this discovery).

The only two electromagnetic invariants being important in the
Einstein--Maxwell theory can be easily introduced: \be \label{I1}
I_1=-2\ast(F\wedge\ast F)=F_{\mu\nu}F^{\mu\nu}=2
\left(\mathbf{B}^2-\mathbf{E}^2 \right), \ee \be \label{I2}
I_2=2\ast(F\wedge F)=F\!\stackrel{ \textnormal{\small$\ast$}}{
\textnormal{\scriptsize$\mu\nu$}} F^{\mu\nu}=4 \textnormal{{\bf
E}}\bullet\textnormal{{\bf B}}. \phantom{aaaaaa.} \ee These
invariants enter the following important identities: \be
\label{crafty} F_{\mu\nu}F^{\lambda\nu}-F\!\stackrel{
\textnormal{\small$\ast$}}{\textnormal{\scriptsize$\mu\nu$}}
F\!\stackrel{ \textnormal{\scriptsize$\lambda\nu$}}{
\textnormal{\small$\ast$}}=\frac{1}{2}I_1\,\delta^\lambda_\mu, ~ ~
~ F\!\stackrel{\textnormal{\small$\ast$}}{
\textnormal{\scriptsize$\mu\nu$}}F^{\lambda\nu}=\frac{1}{4}I_2
\,\delta^\lambda_\mu. \ee In fact, $I_2$ is an axial (pseudo-)
invariant whose square behaves as a usual scalar.

The electromagnetic stress-energy tensor is \cite{Mitsk58,Mitsk06}
\be \label{Tmunu}
T^\nu_\mu=\frac{1}{4\pi}\left(\frac{1}{4}F_{\kappa\lambda}
F^{\kappa\lambda} \delta^\nu_\mu-F_{\mu\lambda}F^{\nu\lambda}
\right)=-\frac{1}{8\pi}\left( F_{\mu\lambda}F^{\nu\lambda}+F\!
\stackrel{\textnormal{\small$\ast $}}{\textnormal{\scriptsize$\mu
\lambda$}} F\!\stackrel{ \textnormal{\scriptsize$\nu\lambda
$}}{\textnormal{\small$\ast$}}\right) \ee (in Gaussian units). Its
(single) contraction with arbitrary monad includes the
electromagnetic energy density and Poynting vector in that frame,
\be \label{Ttau} T^\nu_\mu \tau_\nu=\frac{1}{8\pi}\left[\left
(\textnormal{{\bf E}}^2+ \textnormal{{\bf B}}^2\right)\tau_\mu+2
(\textnormal{{\bf E}} \times\textnormal{{\bf B}})_\mu\right], \ee
and the squared expression is (see (\ref{crafty}) and {\it cf.}
\cite{Whee,Rain1})
\begin{multline} \label{Ttausq} T^\nu_\mu T^\mu_\xi\tau_\nu
\tau^\xi=\frac{1}{(8\pi)^2}\left[\left(\textnormal{{\bf E}}^2+
\textnormal{{\bf B}}^2\right)^2-4(\textnormal{{\bf E}}\times
\textnormal{{\bf B}})^2\right] \\
\equiv\frac{1}{(8\pi)^2}\left[\left(\textnormal{{\bf B}}^2-
\textnormal{{\bf E}}^2\right)^2+4(\textnormal{{\bf E}}\bullet
\textnormal{{\bf B}})^2\right]=\frac{1}{(16\pi)^2}\left({I_1}^2+
{I_2}^2\right). \end{multline} It is interesting that these
constructions are not only scalars under transformations of
coordinates, but they are also independent of the reference frame
choice: the right-hand side does not involve any mention of the
monad at all.

\section{A classification of electromagnetic fields} \label{s4}
\setcounter{equation}{0}

The simple and exhaustive classification of electromagnetic fields
is based on existence of only two invariants, (\ref{I1}) and
(\ref{I2}), built with the field tensor $F_{\mu\nu}$, while all
other invariants are merely algebraic functions of these two
invariants (if not vanish identically). Since $I_2$ itself is a
pseudo-invariant (axial scalar) which acquires the factor
sign$(J):=J/|J|$ under a general transformation of coordinates,
$J$ being its Jacobian, the concrete sign of $I_2$ does not matter
in our classification.

In terms of $I_1$ the invariant classification suggests three
types of fields: $I_1<0$ is the electric type (the electric field
dominates), $I_1>0$ gives the magnetic type, and to $I_1=0$, the
null type corresponds. The pseudo-invariant $I_2$ permits to work
out the classification in more detail: we get additional subtypes,
impure ($I_2\neq 0$) and pure ($I_2=0$).

Below we shall see how this classification enables us to find
reference frames most adequately suitable for description of
concrete electromagnetic fields and even to construct new exact
solutions of Einstein--Maxwell's equations. It also gives a
natural base for straightforward physical interpretation of these
fields.

\section{Propagation of electromagnetic fields} \label{s5}
\setcounter{equation}{0}

Considering the propagation of electromagnetic field, we do not
include the high-frequency limits related to field discontinuities
(bicharacteristics). The Poynting vector plays an important r\^ole
in electrodynamics having two distinct meanings: of the energy
density flow and of the linear momentum density due to symmetry of
the electromagnetic energy-momentum tensor (in natural units
velocity is dimensionless and that of light in a vacuum is $c=1$).
It is worth giving more comments on physical interpretation of the
Poynting vector. It does not always describe propagation of
extractable energy of the field and even a real motion (see also
\cite{SyngeHerm}); the exclusion is here related to the special
case of static and stationary fields (whose frequency is equal to
zero). Thus the Poynting vector, together with the electromagnetic
energy density, determines (sometimes formally) the propagation
three-velocity of electromagnetic field with respect to the
reference frame in which the expression (\ref{Ttau}) is given. We
take this velocity according to Landau and Lifshitz \cite{LanLif}
(see the problem in p. 69) as \be \label{emprop} \frac{\mathbf{v}
}{1+\mathbf{v}^2}=\frac{\mathbf{E}\times\mathbf{B} }{\mathbf{E}^2
+\mathbf{B}^2} \ee (an alternative definition see in \cite{Pauli},
p. 115, formula (312), \be \label{Pauli} \mathbf{v}=2
\frac{\mathbf{E}\times\mathbf{B} }{\mathbf{E}^2 +\mathbf{B}^2},
\ee but this definition is false as it can be seen from subsection
\ref{s11.2} below). In the preceding pages in \cite{LanLif}, an
interesting discussion of electromagnetic invariants is worth
being noted. From (\ref{Ttausq}) and (\ref{emprop}) we see that
\be \label{modv} 0\leq\frac{|\mathbf{v}|}{1+\mathbf{v}^2}=
\frac{1}{2} \sqrt{1-\frac{{I_1}^2+{I_2}^2}{4(\mathbf{E}^2+
\mathbf{B}^2)^2}}=\frac{|\mathbf{E}||\mathbf{B}|}{\mathbf{E}^2+
\mathbf{B}^2}|\sin\psi|\leq\frac{1}{2}, \ee $\psi$ being the angle
between {\bf E} and {\bf B} in the strict local Euclidean sense;
moreover, the function $|\mathbf{v}|/(1+\mathbf{v}^2)$ is
everywhere monotonic. In particular, this means that the
propagation of all pure null fields ($|\mathbf{E}|=|\mathbf{B}|$,
$\psi=\pi/2$) occurs with the unit absolute value of the
three-velocity, the velocity of light, and all other
electromagnetic fields propagate with sub-luminal velocities which
can always be made equal to zero in corresponding co-moving
reference frames. This is the general-relativistic conclusion,
only expressed in three-dimensional notations characteristic to
the general reference frame theory.

\section{Dealing with pure electromagnetic fields} \label{s6}
\setcounter{equation}{0}

Pure electromagnetic fields represent the simplest cases,
especially in the non-null types when there always exist reference
frames in which either magnetic or electric field can be easily
eliminated. The pure null type requires more thorough examination
involving a consideration of the Doppler effect (here, its
generalized counterpart) which we have to discuss below in more
detail.

\subsection{Pure electric and magnetic type fields} \label{s6.1}

Vanishing of the second invariant, $I_2$, means that the
electromagnetic field tensor, or its dual conjugate, is a simple
bivector in all reference frames (the second of two necessary and
sufficient conditions is four-dimensionality of the manifold under
consideration), thus \be \label{simbiv} F=U \wedge V \textnormal{
or } \ast F=P\wedge Q, \ee $U$, $V$, $P$, and $Q$ being
four-(co)vectors. In the first case, \be \label{FF} I_1=2\left((U
\cdot U)(V\cdot V)-(U\cdot V)^2\right) \ee obviously is negative
if one of these vectors is timelike (say, $U$) and another,
spacelike ($V$), thus $F$ will pertain to the pure electric type
(or, similarly, for $\ast F$, to the pure magnetic type; see also
an alternative case considered in subsection \ref{s11.1} when
vector $U=R$ is null). Normalizing timelike $U$ to unity (the
extra coefficient may be included in $V$), we can take the
normalized $U$ as a new monad in the choice of reference frame and
immediately see that in this frame the magnetic vector
automatically vanishes. It remains only to show that our
supposition (timelike $U$ and spacelike $V$) is sufficiently
general; this can be easily proven using the substitution
$V\Longrightarrow V+aU$ which does not change $F$. Similarly we
treat the problem of eliminating the electric field in the pure
magnetic case using $\ast F$ in (\ref{simbiv}).

\subsection{Pure null type fields and the Doppler effect}
\label{s6.2}

Pure null type fields have both invariants equal to zero, but the
very fields remain non-trivial in any non-degenerate system of
coordinates as well as in any realistic reference frame (here, in
the sense of {\bf E} and {\bf B}, simultaneously), although these
three-vectors do transform under changes of reference frames, and
their components transform under transformations of coordinates.
The monad $\tau$ under these transformation should remain always
timelike, and the Jacobian of the transformation of coordinates
has to be non-zero and non-$\infty$.

As an example we consider in this subsection a
special-relativistic plane electromagnetic wave in a vacuum ($k=
\omega$) written in Cartesian coordinates, \be \label{emw}
\mathbf{E}=\{0,E \cos[\omega(x-t)],0\}, ~ ~
\mathbf{B}=\{0,0,E\cos[\omega(x-t)]\}, \ee and apply to it the
Lorentz transformation $t,\mathbf{r} \Rightarrow t',\mathbf{r}'$
with the three-velocity $\pm v$ in the positive/negative direction
of the $x$ axis using the well-known change of {\bf E} and {\bf B}
under this transformation. The resulting electromagnetic field
then is \be \label{emwnew}
\mathbf{E}'=\{0,E'\cos[\omega'(x'-t')],0\}, ~ ~
\mathbf{B}'=\{0,0,E'\cos[\omega' (x'-t')]\} \ee where \be
\label{Dopp} E'= \sqrt{\frac{1\mp v}{1\pm v}}E, ~ ~
\omega'=\sqrt{\frac{1\mp v}{1\pm v}}\omega. \ee It is clear that
the expression of $\omega'$ in (\ref{Dopp}) describes the
longitudinal Doppler effect while $E'$ gives the accompanying
change of the wave intensity. Since the latter is an integral part
of the longitudinal Doppler effect, we consider the complete
expression (\ref{Dopp}) as its natural generalization; the
description of transversal Doppler effect has to be generalized in
the similar way.

It seems that this generalization of the Doppler effect is not
encountered in physics textbooks. Nevertheless, it is generally
used as an important hint in the interpretation of the well-known
astrophysical phenomenon of ultrarelativistic particles' jet pairs
emitted by cores of some galaxies (the jet moving away from the
observer not only has lower frequency, but also correspondingly
lower intensity, thus this jet sometimes escapes to be observed).
There is also a static ($\omega =0$) particular case of pure null
electromagnetic fields involving mutually orthogonal constant
vectors {\bf E} and {\bf B} (let us call it ``Cartesian case''
whose cylindrically symmetric analogue is used in some experiments
involving electromagnetic fields with non-zero angular momentum
without a genuine rotation). Such Cartesian pure null fields
manifest only intensity part of the Doppler effect since in this
case $\omega=0=\omega'$ in (\ref{Dopp}). Thus the pure null
electromagnetic fields can be adjusted to any non-zero and
non-$\infty$ values of their intensity and frequency (only the
relation of frequency to intensity remaining constant, and if the
frequency had not been equal to zero from the very beginning). A
complete transformation away of initially non-trivial pure null
fields is however impossible in any non-degenerate frame,
representing only asymptotic and not real possibility. We
considered above the case of a plane-polarized wave, but similar
approach works in the circular polarization case as well.

\section{Duality rotation and electromagnetic fields with the
same spacetime geometry} \label{s7} \setcounter{equation}{0}

Let us introduce a new electromagnetic field tensor (2-form) \be
\label{calF} \mathcal{F}=(k+l\ast)F, ~ ~ \ast\mathcal{F}=(k\ast-
l)F \ee ($\ast$ is the Hodge star), where $F$ is the
electromagnetic field tensor belonging to some given exact
self-consistent Einstein--Maxwell solution, $k$ and $l$ being some
scalar functions to be further determined. We now set the
condition that the new field $\mathcal{F}$ has to produce the same
energy-momentum tensor which follows from the old field $F$. Since
geometry is well determined by the energy-momentum tensor, from
the Bianchi identities it then follows that the standard general
relativistic Maxwell equations for both fields, old and new, will
be equally satisfied if the old field has no electromagnetic
sources, or the sources are localized at the singularity of the
old and new fields where the standard classical theory is not
applicable.

The calligraphic letters will be used for all concomitants of the
new electromagnetic field. Thus \be \label{calF1} \mathcal{F}_{\mu
\nu}=kF_{\mu\nu}+lF\!\!\stackrel{\textnormal{\small$\ast$}
}{\textnormal{\scriptsize$\mu\nu$}}, ~ ~ \ast\mathcal{F}=k
F\!\!\stackrel{\textnormal{\small$\ast$} }{\textnormal{\scriptsize
$\mu\nu$}}- lF_{\mu\nu}, \ee  \be \label{calEB} \mathcal{E}=
k\mathbf{E}-l\mathbf{B}, ~ ~ ~ \mathcal{B}=k\mathbf{B}+l
\mathbf{E}, \ee
$$
\mathcal{I}_1=\mathcal{F}_{\mu\nu }\mathcal{F}^{\mu\nu }=(k^2-
l^2)I_1+2klI_2, ~ ~ \mathcal{I}_2=\mathcal{F}\!\!\stackrel{
\textnormal{\small$\ast$} }{\textnormal{\scriptsize$\mu\nu$}}\!
\mathcal{F}^{\mu\nu}=(k^2-l^2 )I_2-2klI_1.
$$
A simple calculation yields $\mathcal{T}^\mu_\nu=
\frac{k^2+l^2}{4\pi} \left( \frac{1}{4}I_1\delta^\mu_\nu-F^{
\alpha\mu}F_{\alpha\nu}\right)=(k^2+l^2)T^\mu_\nu$ (the terms with
$I_2$ cancel automatically for arbitrary $I_2$). Hence the
coincidence of geometries created by the two fields is guaranteed
iff \be \label{lk} k=\cos\alpha, ~ ~ l=\sin\alpha, \ee so that \be
\label{calI} \mathcal{I}_1=\cos2 \alpha\,I_1+\sin2\alpha\, I_2, ~
~ \mathcal{I}_2=\cos2\alpha\,I_2-\sin2\alpha\,I_1. \ee

Duality rotation, how the ``transformation'' (\ref{calF}) with
(\ref{lk}) is now interpreted (see \cite{MisWh}), does not change
the 4-geometry compatible with the new electromagnetic field. This
geometry remains the same as that created by the old field (see
also \cite{Rain1,MisWh,Whee}). The ``angle'' (complexion) $\alpha$
of the duality rotation is, of course, an axial scalar function of
coordinates which we shall now concretely determine.

\section{Construction of new Einstein--Maxwell solutions {\it
via} duality rotation} \label{s8} \setcounter{equation}{0}

First, let us see how restrictive is the duality rotation. The
relations (\ref{calI}) lead to a general conclusion \be
\label{Pyth} {\mathcal{I}_1}^2+ {\mathcal{I}_2}^2={I_1}^2+{I_2}^2
\ee from where it follows that the pure null property is invariant
under the duality rotation and it is impossible to obtain from any
other type a pure null solution using this method. Together with
the considerations of subsection \ref{s6.2}, this means that pure
null fields sharply differ from all other electromagnetic fields.

Now, let us see if pure subtypes ($\mathcal{I}_2=0 ~ \Rightarrow ~
{\mathcal{I}_1}^2={I_1}^2+{I_2}^2$) can be obtained from the
impure fields ($I_2\neq 0$). The second relation in (\ref{calI})
then yields \be \label{Ialph1} \cot2\alpha=I_1/ I_2 \ee which with
the first relation gives \be \label{pureI} \mathcal{I}_1=\frac{I_2
}{\sin2\alpha}= \frac{I_1}{\cos2\alpha} =\cos2\alpha\frac{{I_1}^2
+{I_2}^2}{I_1}= \sin2\alpha\frac{{I_1}^2+{I_2}^2}{I_2}. \ee (In
fact, we have to perform straightforward calculations for every
concrete solution $F$ and see if $I_2$ would be sign-definitive or
not in the desired region. Though the last possibility seems to be
excluded by our initial supposition, it could be, naturally,
softened: the duality rotation should reduce to the identity
transformation at {\it loci} where $I_2$ becomes equal to zero.)
From (\ref{Ialph1}) and (\ref{pureI}) we come to the following
conclusions: if the new field has to be pure electric
($\mathcal{I}_1<0$), while $I_1<0$ and $I_2>0$, the ``angle''
$\alpha$ has to be such that $\sin2\alpha<0$ and $\cos2\alpha>0$;
for $I_1<0$ and $I_2<0$, $\alpha$ has to give $\sin2\alpha>0$ and
$\cos2\alpha>0$; for $I_1>0$ and $I_2>0$, there has to be $\sin2
\alpha<0$ and $\cos2\alpha<0$; for $I_1>0$ and $I_2<0$, $\sin2
\alpha>0$ and $\cos2\alpha<0$. Similarly, we determine the
position of $\alpha$ for the pure magnetic new field.

For the null (now impure) type of the new field ($\mathcal{I}_1=0
~ \Rightarrow ~ {\mathcal{I}_2}^2={I_1}^2+{I_2}^2$) we have to use
the first relation in (\ref{calI}) yielding \be \label{Ialph2}
\tan2 \alpha=-I_1/I_2. \ee The second relation gives \be
\label{impurenull} \mathcal{I}_2=\frac{I_2}{\cos2\alpha}=
\frac{I_1}{\sin2\alpha}, ~ ~ etc; \ee the procedure of
determination of the position of $\alpha$ is the same as above,
here only $\mathcal{I}_2\neq 0$, and both signs of $\mathcal{I}_2$
are equally admissible. It is clear that we can perform the
inverse duality rotation in all these cases (in particular, coming
from the impure null to pure electric or magnetic type fields).

Thus the pure and impure electric and magnetic types form together
with the impure null type a mutually ``transformable'' ({\it via}
duality rotation) group of electromagnetic fields disconnected
from the pure null type.

\section{Impure electromagnetic fields: parallelizing of
{\bf E} and {\bf B}} \label{s9} \setcounter{equation}{0}

In this section we again use the classification of electromagnetic
fields in two senses: in the proper one, {\it i.e.} with respect
to $F$, and, simultaneously, in the sense of the new field
$\mathcal{F}$ introduced in (\ref{calF}), but we now look for
information received from $\mathcal{F}$ about the old field $F$.
It is already clear that when $F$ is impure (electric, magnetic,
or null), the field $\mathcal{F}$ can be chosen as pure (electric
or magnetic in everyone of these cases). An interesting feature
here is that the reference frame in which only one field (electric
or magnetic in the sense of $\mathcal{F}$) survives, is precisely
that in which {\bf E} and {\bf B} following from $F$ are mutually
parallel, and the parallelization procedure becomes completely
reduced just to determination of this canonical frame (say,
$\tau'$). Thus, while the relations \be \label{pure}
\mathcal{E}\bullet \mathcal{B}\equiv
\mathcal{E}'\bullet\mathcal{B}'=0 \ee are frame-invariant (the
field $\mathcal{F}$ is chosen as belonging to the pure subtype),
the property $\mathbf{E}'\parallel\mathbf{B}'$ is realized only in
the canonical frame where either $\mathcal{E}'$ or $\mathcal{B}'$
vanishes.

The field $\mathcal{F}$ may be considered as a merely auxiliary
one (essentially, in special relativity where its r\^ole in
generation of gravitational field is neglected), so that the
parallelization procedure then may be managed even without the use
of the strict duality rotation with the ``angle'' $\alpha$ and the
relation (\ref{Pyth}), but when we simply take, {\it e.g.}, \be
\label{calFk} \mathcal{F}=(1+k\ast)F. \ee The calculations
following from this ansatz are simple, but somewhat cumbersome,
and we omit them, especially since they will not be used in this
paper.

\section{Examples in general relativity} \label{s10}
\setcounter{equation}{0}

In this section we consider two particular electromagnetic fields
self-consis\-tent\-ly sharing one and the same four-dimensional
geometry: in the first subsection, the standard Kerr--Newman (KN)
rotating charged black hole, and in the subsection \ref{s10.2},
its generalizations to the black holes created by specific
mixtures of electric charge and magnetic monopole distributions
rotating as the KN singular ring. The first example corresponds to
an impure electromagnetic field, while the next three ones belong
to the pure electric, pure magnetic and impure null types, thus
representing new black-hole exact solutions of Einstein--Maxwell
equations. The first two new solutions admit reference frames in
which there is no magnetic or no electric fields in the whole
spacetime.

\subsection{The Kerr--Newman solution} \label{s10.1}

The KN metric tensor is taken in the Boyer--Lindquist (BL)
coordinates as \be \label{KN-BL} \begin{array}{l} ds^2=
\displaystyle\frac{\Delta}{\rho^2}
\left(dt-a\sin^2\vartheta\,d\varphi\right)^2-\frac{\rho^2}{
\Delta}dr^2-\rho^2d\vartheta^2\\
\phantom{ds^2}-\displaystyle\frac{\sin^2\vartheta}{
\rho^2}\left[(r^2+a^2)d\varphi-adt\right]^2 \end{array} \ee where
$\rho^2=r^2+a^2\cos^2\vartheta$, $\Delta=r^2-2Mr+Q^2+ a^2$. Thus
the orthonormal 1-form basis outside the singularity $\rho=0$
reads \be \label{11.2} \left.
\begin{array}{ll} \hspace{-7.pt}\theta^{(0)}=
\displaystyle\frac{\sqrt{\Delta}}{\rho}\left(dt-a\sin^2\vartheta\,
d\varphi\right), & \theta^{(1)}=\displaystyle\frac{\rho}{\sqrt{
\Delta}}dr,\\ \hspace{-7.pt}\theta^{(2)}=\rho d\vartheta, &
\theta^{(3)}= \displaystyle
\frac{\sin\vartheta}{\rho}\left[(r^2+a^2) d\varphi-adt\right],
\end{array}\right\} \ee so that $d\varphi=\frac{a}{\rho\sqrt{\Delta
}}\theta^{(0)}+\frac{1}{\rho\sin\vartheta}\theta^{(3)}$. Further,
the electromagnetic 1-form four-potential ({\it cf.} the usual
Coulomb potential) and the field tensor (2-form) are \be
\label{11.3} \left.\begin{array}{l} A\equiv A_{(\alpha)}
\theta^{(\alpha)}= \displaystyle\frac{Qr}{\rho\sqrt{\Delta}}
\theta^{(0)} ~ ~ \textnormal{ and}\\ ~ \\
F\equiv\frac{1}{2}F_{(\alpha)(\beta)} \theta^{(\alpha)}
\wedge\theta^{(\beta)}=dA \\ \phantom{F}= \displaystyle\frac{
Q}{\rho^4}\left[(r^2-a^2 \cos^2\vartheta) \theta^{(0)}\wedge
\theta^{(1)}-2ar\cos\vartheta\,\theta^{(2)} \wedge\theta^{(3)}
\right], \end{array}\right\} \ee respectively (see for some
details \cite{DKSch,HEL,tHooft,Kerr,KSMH,Exact2,Teuk,Twns}). Since
\be \label{propbeta} 4a^2r^2\cos^2\vartheta+\left(r^2- a^2\cos^2
\vartheta\right)^2=\rho^4 \ee (this confirms the presence in $F$
of the factor $r^2-a^2\cos^2\vartheta$ which first could seem to
be somewhat unnatural), we can now introduce an ``angle'' $\beta$
as \be \label{beta} \sin\beta=\frac{2ar\cos \vartheta}{\rho^2}, ~
~ \cos\beta=\frac{r^2-a^2\cos^2\vartheta}{ \rho^2}. \ee Then the
electromagnetic field $F$ reads \be \label{FasDR}
F=\frac{Q}{\rho^2}(\cos\beta+\sin\beta\ast)\left(\theta^{(0)}
\wedge\theta^{(1)}\right), \ee obviously involving a duality
rotation, and the electromagnetic invariants read \be \label{11.6}
I_1=-\frac{2Q^2 }{\rho^4}\cos2\beta, ~ ~ I_2=\frac{2Q^2}{\rho^4}
\sin2\beta, \ee so that the construction invariant under duality
rotation (\ref{Pyth}) in the KN solution case is ({\it cf.} the
Coulomb and Reissner--Nordstr\"om fields where, of course, $a=0$)
\be \label{I12sqr} {I_1}^2+{I_2}^2= \frac{4Q^4}{\rho^8}. \ee

From (\ref{11.6}) and (\ref{beta}) we immediately find that on the
``plane'' $\cos\vartheta=0$ and the ``sphere'' $r=0$ (it is well
known that the ``negative region of space'' with $r<0$ makes a
certain sense in this spacetime) the invariants take values
$$
I_1=-\frac{2Q^2}{r^4}, ~ ~ I_2=0 ~ \textnormal{ and } ~ I_1=-
\frac{2Q^2 }{a^4\cos^4\vartheta}, ~ ~ I_2=0,
$$
respectively, thus the field $F$ belongs there to the pure
electric type (in the BL frame, the magnetic field is already
eliminated in this region). The intersection of these 2-surfaces
is the well-known singular rotating Kerr ring where $I_2$ vanishes
when we approach to the ring from these mutually orthogonal
directions. This would be in conformity with the usual
interpretation of the ring as rotating with the velocity of light
if that $I_1$ also tended there to zero, though this is not quite
the case. The vanishing of $I_1$ occurs only in the limits in four
directions along which $r=\epsilon(3\pm2\sqrt{2})a^2\cos^2
\vartheta$ where $\epsilon=\pm 1$ (without admission to
simultaneously take only both similar signs in the whole formula).
Further, when $r^2=a^2 \cos^2\vartheta$, we have the pure magnetic
type field (with already eliminated electric field in the BL
frame) with $I_1=\frac{3Q^2}{4a^4\cos^4\vartheta}=
\frac{3Q^2}{4r^4}$, thus, if we come to the ring from the
corresponding directions, the field will be purely magnetic. The
electromagnetic field around the Kerr ring in KN solution is in
fact very diverse, like a patchwork quilt.

To find how behaves the propagation velocity {\bf v} of this
field, we have to calculate its energy density and Poynting
vector, but let us begin with {\bf E} and {\bf B} in the BL frame
$\tau=\theta^{(0)}$, see (\ref{11.2}). Already before any
calculations, only looking at the definitions (\ref{elE}) and
(\ref{magB}), one understands that these two vectors and
$\theta^{(1)}$ are everywhere collinear, so that the Poynting
vector identically vanishes, as well as {\bf v} does (this is
natural, since both metric coefficients and the field tensor
components do not depend on $t$ and $\varphi$, and we already {\it
are} in the KN-field's co-moving frame). The pure subtype can be
realized (if we suppress duality rotations) only due to local
vanishing of {\bf E} or {\bf B} in the BL frame. In this co-moving
frame the electromagnetic energy density is everywhere equal to
\be \label{KNw}
T_\mu^\nu\tau^\mu\tau_\nu=\frac{1}{8\pi}(\mathbf{E}^2+\mathbf{B}^2
)=\frac{1}{16\pi}\sqrt{{I_1}^2+{I_2}^2}=\frac{Q^2}{8\pi\rho^4},
\ee see (\ref{modv}).

We have found here that the KN solution is ``anisotropic'' and
``inhomogeneous'' in the sense of distribution of electric and
magnetic types of the $F$ field, although three-vector fields {\bf
E} and {\bf B} are always collinear in the spacetime of KN black
hole in the BL frame, but there are surfaces on which either
magnetic or electric field vanishes. In order to better understand
if this structure of electromagnetic field belonging to the KN
solution is inevitable (however, see also \cite{HEL}), we shall
further apply to the KN solution the method developed in sections
\ref{s7} and \ref{s8} (as a broadening and modernization of the
Rainich--Misner--Wheeler duality rotation approach). We shall find
that it is easy to radically modify the KN solution obtaining
rotating black holes with electromagnetic fields of everywhere
pure electric or pure magnetic, or impure null type.

\subsection{``New'' rotating black hole solutions with pure
elec\-tric, pure mag\-ne\-tic, and impure null fields $F$ in KN
geometry} \label{s10.2}

Here we apply duality rotation \be \label{dualrot}
\mathcal{F}=(\cos\alpha+\sin\alpha\ast)F, ~ \textnormal{or} ~
\ast\mathcal{F}=(\cos\alpha\ast-\sin\alpha)F \ee [the combination
of (\ref{calF}) and (\ref{lk})] to the KN electromagnetic field
(\ref{11.3}). Since our aim is to construct a pure subtype field
($\mathcal{I}_2=0$), we already have from (\ref{Pyth}) ${\mathcal{
I}_1}^2={I_1}^2+{I_2}^2$ while the sign of $\mathcal{ I}_1$ is
determined by the choice of position of $\alpha$ in the angle
diagram and (\ref{Ialph1}). Taking the expression (\ref{FasDR})
and using the obvious algebra of duality rotation, we rewrite the
expression (\ref{dualrot}) as \be \label{2drot} \mathcal{F}=
\frac{Q}{\rho^2}\left[\cos(\alpha+\beta)+\sin(\alpha+\beta)\ast
\right]\left(\theta^{(0)}\wedge\theta^{(1)}\right). \ee Now,
putting $\alpha+\beta=0$, we immediately come to the pure electric
field $\mathcal{F}_{\mathrm{el}}$ \be \label{pureelKN}
\mathcal{F}_{\mathrm{el}}=\frac{Q}{\rho^2}\theta^{(0)}\wedge
\theta^{(1)} \ee generating the same KN geometry which everybody
associates with the KN ``patchwork'' electromagnetic field $F$
(\ref{11.3}). The pure magnetic field \be \label{puremagKN}
\mathcal{F}_{\mathrm{mag}}=\frac{Q}{\rho^2}
\theta^{(2)}\wedge\theta^{(3)} \ee is similarly obtained with the
use of $\alpha+\beta=-\pi /2$.

Finally, we have to add the third new case of KN-like black holes,
those with null type electromagnetic field when in (\ref{2drot})
$\alpha+\beta=\pi/4+n\pi/2$ is taken (naturally, this field now
belongs to the impure subtype). With $n=0$, it reads \be
\label{nullKN} \mathcal{F}_{\mathrm{null}}= \frac{Q}{\sqrt{2}
\rho^2} \left(\theta^{(0)}\wedge\theta^{(1)}-\theta^{(2)}\wedge
\theta^{(3)}\right). \ee

The (contracted) Bianchi identities guarantee satisfaction of
Maxwell's equations outside the ring singularity for the new field
$\mathcal{F}$ in this geometry, while the presence of magnetic
monopole distribution existing here only on the singular Kerr ring
should not create any problem since at the singularity the
classical laws of physics obviously fail to work. One may say that
the magnetic monopole distribution (as well as that of the
electric charge) can, as this is shown above, exactly compensate
the magnetic (electric) field created by a rotating charge
(rotating monopole) distribution, but this is not precisely the
case. In fact, we encounter in this situation a more complicated
superposition of dynamical and kinematic effects, since the ($\tau
=\theta^{(0)}$)-reference frame is not an inertial one: it
involves both acceleration and rotation (the latter is present due
to $\theta^{(0)}\wedge d\theta^{(0)}\neq 0$, see the definitions
in \cite{Mitsk06}). In the same ref. 11 (chapter 4, pp. 86, 87,
90, and 91) it is shown that in the classical Maxwell equations
and in laws of motion of electric charges, both written in
non-inertial reference frames, there appear kinematic terms of the
monopole nature. In equations of motion they bear the name of
kinematic forces (forces of inertia), thus in the field equations
let us speak of kinematic sources. While dynamical force and
source are originated by the same interaction term in the action
integral (only the variational procedure is performed with respect
to particle's world line and to field's potential,
correspondingly), their kinematic counterparts automatically
appear in the respective dynamical equations written (and
experimentally investigable) in non-inertial frames. It is
interesting that kinematic and dynamical counterparts of forces,
as well as of sources, have rather similar structure, despite
their different origin, thus making them recognizable.

\section{Examples in special relativity} \label{s11}
\setcounter{equation}{0}

\subsection{Li\'enard--Wiechert's field: the pure electric type}
\label{s11.1}

The Li\'enard--Wiechert (LW) field is special relativistic
electromagnetic field generated by an arbitrarily moving electric
charge $Q$ (we restrict our consideration to an arbitrary timelike
world line of the charge). See the details of deduction of this
field in \cite{Mitsk05} where we used the future light cone (the
case of retarded field, precisely like in the present paper); an
arbitrary mixture of retarded and advanced fields can be found in
\cite{SyngeSR}. Thus the retarded point on the charge world line
$x'^\mu$ is connected with the four-dimensional point $x^\mu$
(where the field is determined) by the null vector $R^\mu=x^\mu-
x'^\mu$ (we choose for simplicity the Cartesian coordinates in
this special relativistic treatment). The four-potential then
reads \be \label{LW} A^\mu= \frac{Qu'^\mu}{D} \ee where $u'^\mu=
dx'^\mu/ds'$ ($u'\cdot u'=1$) is the retarded four-velocity of the
charge and $D= u'\cdot R\equiv\sqrt{-\mathbf{D}^\mu
\mathbf{D}_\mu}$, while $\mathbf{D}^\mu=R^\nu b^\mu_\nu$,
$b=g-u'\otimes u'$ simultaneously being the projector on local
retarded three-dimensional subspace orthogonal to $u'$ and the
spatial three-metric on this subspace (with the signature
$0,-,-,-$). The retarded four-acceleration of the charge is
$a'^\mu=du'^\mu/ds'$ (naturally, $a'\cdot u'\equiv 0$). A simple
calculation yields the LW field tensor \be \label{FTLW} F=R\wedge
V, ~ ~ V=\frac{Q}{D^2} \left(a'+u'\frac{1-a'\cdot R}{D} \right)
\ee (the second field invariant $I_2$ automatically vanishes). The
first invariant is \be \label{1Inv} I_1=-\frac{2 Q^2}{D^4}<0 \ee
(remarkably, its structure is exactly Coulombian). Thus the LW
field pertains to the pure electric type everywhere outside the
point charge's world line.

Combining $R$ with $V$ in (\ref{FTLW}), one can change the null
vector $R$ to a timelike one, $U$, and thus reduce the problem to
that discussed in subsection \ref{s6.1}. However it is much
simpler to (algebraically) regauge the vector $V ~ \rightarrow ~
W=V+\frac{Q}{D^2}lR$ (the fractional coefficient is put only for
convenience, and $l$ is a scalar function still to be determined);
this does neither change the field tensor, $F=R\wedge W$, nor
produce any $l^2$-term in further calculations. Applying now the
1-form definition of the magnetic vector in a $\tau$-frame
(\ref{magB}) and taking the monad as $\tau=NW$ where the scalar
normalization factor is $N=(W\cdot W)^{-1/2}$, we come to
$\mathbf{B}=0$ in this frame. The problem is thus reduced to a
proper choice of $l$ such that $W$ will be a suitable real
timelike vector. We see that \be \label{V1} W=\frac{Q}{D^2}\left(
a'+\frac{1-a'\!\cdot\! R}{D}\,u' +lR\right). \ee Then its square
takes an unexpectedly simple form \be W\!\cdot\!
W=\left(\frac{Q}{D^2} \right)^2\left[ a'\!\cdot\! a'+
\frac{(1-a'\!\cdot\! R)^2}{D^2}+2l\right]. \ee In fact, $l$ still
remains arbitrary (this means that there is a continuum of such
different co-moving frames). Let it be \be l=
\frac{1}{2}\left[\frac{1}{D^2}-a'\!\cdot\! a'-\frac{(1-a'\!\cdot\!
R)^2}{D^2}\right] \ee (the first term in the square brackets,
$1/D^2$, got its denominator to fit the dimensional
considerations). Finally, $W\cdot W=\left(
\frac{Q}{D^3}\right)^2>0$ and \be \label{taucomov} \tau=Da'+
\left(1-a'\!\cdot\! R\right)u'+\frac{1}{2D}\left[1-D^2a'\! \cdot\!
a'-\left(1-a'\!\cdot\! R\right)^2\right]R \ee (it is clear that
$\tau\cdot\tau=+1$). By its definition, the monad $\tau$ describes
the reference frame co-moving with the LW electromagnetic field:
in this frame the Poynting vector of the field vanishes, the
electromagnetic energy flux ceases to exist due to the absence of
magnetic part $\overset{\textnormal{\tiny $\tau$}}{\mathbf{B}}$ of
the field in this frame (applicable at any finite distance $D$,
not asymptotically), and $F$ can be rewritten as
$F=\frac{Q}{D^3}R\wedge\tau$. The expression (\ref{elE}) now
yields \be \label{Ecomov} \overset{\textnormal{\tiny
$\tau$}}{\mathbf{E}}=\ast(\tau\wedge\ast F)=
\frac{Q}{D^3}\ast[\tau\wedge\ast
(R\wedge\tau)]=\frac{Q}{D^2}\overset{\textnormal{\tiny
$\tau$}}{\mathbf{n}} \ee which is, up to an understandable
reinterpretation of notations, exactly the form known as the
Coulomb field vector. Here the unitary vector
$\overset{\textnormal{\tiny $\tau$}}{\mathbf{n}}=
\overset{\textnormal{\tiny $\tau$}}{\mathbf{D}}/D$ is normal to
the $\tau$-congruence, while $R\cdot u'=
\overset{\textnormal{\;\tiny $u'$}}{D}=D=
\overset{\textnormal{\;\tiny $\tau$}}{D}=R\cdot\tau$ and
$\overset{\textnormal{\tiny $\tau$}}{\mathbf{D}}^\mu=b^\mu_\nu
R^\nu$ with $b^\mu_\nu=\delta^\mu_\nu- \tau^\mu\tau_\nu$, hence in
the frame co-moving with the LW field (the reader may choose other
co-moving reference frames taking different $l$s, but our choice
seems to be one of the simplest ones) \be \label{Dhat}
\overset{\textnormal{\tiny $\tau$}}{\mathbf{D}}=-D^2a'-
D\left(1-a'\! \cdot\! R\right)u'+\frac{1}{2}\left[1+D^2
a'\!\cdot\! a'+ \left(1-a'\!\cdot\! R\right)^2\right]R, \ee so
that this $\overset{\textnormal{\tiny $\tau$}}{\mathbf{D}} \neq
\overset{\textnormal{\tiny $u'$}}{\mathbf{D}}$ in fact given after
the formula (11.1) as $\mathbf{D}$ and pertaining to another frame
(co-moving with the retarded charge, not with its field, see for
details \cite{Mitsk05}).

The situation discovered in this subsection can be formulated in a
short and exact form as existence in all spacetime outside the
world line of the charge generating LW's field, of a reference
frame co-moving with this field, {\it i.e.}, a frame in which the
Poynting vector vanishes in all this region (with the exception of
the future null infinity which can be described only
asymptotically, using more topological\footnote{It was a gibe of
the fate with respect to the authors who deliberately disregarded
topology and nevertheless claimed that the LW field contains
electromagnetic radiation, though in this field the Poynting
vector can be easily transformed away by means of a proper choice
of the reference frame in every finite region around the charge's
world line. About the attitude of L.D. Landau and E.M. Lifshitz
toward topology see \cite{Thorne}, pp. 470-471, --- and see also
\cite{LanLif}, p. 173 ff. However, Misner and Wheeler {\it did}
take topology into account in their general-relativistic
considerations \cite{MisWh,Whee}.} than geometrical methods), thus
in this frame there is no flow of electromagnetic energy anywhere.
Of course, this frame is in general a rotating one (see in
\cite{Mitsk05} the expression (4.27) and appendix A), thus the
three-dimensional space is non-holonom (it does not form a global
--- at least, finite --- three-dimensional subspace of the
four-dimensional world; at most, in the presence of rotation there
exist only strictly local (infinitesimal) elements of such a
subspace which do not merge into a finite hypersurface, like
scales of a sick fish in aquarium. Note that this occurs here even
in the special relativity, not only in general theory. Moreover,
the presence of the frame's rotation does not permit
synchronization of clocks being at rest in such a frame. (In the
same spacetime there always exist also an infinite number of
non-rotating frames in which you are welcomed to perform a
synchronization, but in any rotating frame this very procedure is
strictly forbidden. It is curious that while we live all our lives
in our terrestrial rotating frame, its rotation remains
sufficiently slow not to condition us to this non-holonom
psychology.)

Since everywhere outside the LW source (the pointlike charge world
line on which the field singularity occurs) there is no magnetic
field in this frame, and any redistribution of electromagnetic
field cannot take place there, the LW field does not propagate in
this frame, it only can be compressed or rarefied remaining at
rest with respect to the frame in its contraction or expansion,
similar to effects known in relativistic cosmology. This could
seem to be in contradiction with the traditional decomposition of
LW field into the near (induction) and distant (radiation) zones.
The reason for this ``contradiction'' should be seen in the fact
that, although the very Maxwell equations are linear, the physical
characteristics of electromagnetic fields such as their energy
density, Poynting vector (describing, up to a constant factor,
either the energy flow density or linear momentum density of the
field), and stress, are quadratic (or bilinear) in the field
tensor sense. Thus we have not to overlook the ``interaction''
terms between the induction and radiation counterparts in these
characteristics. Note that the elimination of the magnetic part of
LW's field is directly related to its {\it pure} electric type,---
consequently, to the quadratic (bilinear) invariants of this
field. Therefore the ``contradiction'' exists only in a wrong
customary application of the linearity concept to the strictly
nonlinear characteristics even of the electrovacuum
electromagnetic fields. In a certain sense, there should be a way
to reconcile this contradiction considering the asymptotic
behaviour of the field; in any case, this has to correspond to
merely technical details of the problem. Similarly, in his ironic
paper \cite{SyngeHerm}, Synge with his great wit criticized the
existing style of introduction of these same characteristics in
the most widely used textbooks on field theory. Though his
criticism was there somewhat superficial, we find Synge's paper
quite provocative in more profound determination of the quotidian
concepts in our theory using their physical sense.

\subsection{Propagation of a plane electromagnetic wave on the
background of homogeneous magnetic field in a vacuum}
\label{s11.2}

Finally, let us consider a simple, but not yet discussed in
literature problem of electromagnetic waves' propagation in a
time-independent sourceless Maxwell field in a vacuum. For
simplicity, we take the same wave as in subsection \ref{s6.2}, and
the additional Maxwell field is chosen merely as a magnetic one in
the direction of propagation of the wave, with the constant
three-vector~{\bf B}. Obviously, this superposition is an exact
solution of Maxwell's equations, and there cannot be any real
interaction between these two fields since the equations are
linear. We have however already noted that the velocity of
propagation of electromagnetic field is non-linear in terms of
this field's tensor $F$, so that there should, naturally, exist an
observable physical effect in the case of a superposition of such
free Maxwell's fields. There is, of course, an effect which was
already considered and observed in the early history of optics,
that of the standing electromagnetic waves, but nobody still
worried about the seemingly absurd problem formulated above.

Thus we take in Cartesian coordinates $t,x,y,z$ the superposition
of the fields (\ref{emw}) and $\mathbf{B}_{\mathrm H}=Hdx$, {\it
i.e.} \be \label{sup} \mathbf{E}=E\cos[\omega(x-t)]dy, ~ ~
\mathbf{B}=Hdx+E\cos[\omega(x-t)]dz. \ee Obviously, $I_2=0$ due to
the orthogonality of {\bf E} and {\bf B}, and the first invariant
is $I_1=2H^2>0$: this is the pure magnetic type field. The result
of superposition (\ref{sup}) is in fact a specific not precisely
monochromatic wave whose behaviour can be best understood in the
reference frame co-moving with it, and one can find such a frame
using the pure-magnetic property of this wave's field, see
subsection \ref{s6.1}. First, we write the field $\ast F$ [through
(\ref{sup}) in the initial frame $\tau_{\mathrm{in}}=dt$] as a
simple bivector: \begin{align} \label{astFsup} \nonumber \ast F
&=\ast(\mathbf{E} \wedge dt)-\mathbf{B}\wedge dt\\ \nonumber
&=-\left(E\cos[\omega (x-t)]dx\wedge dz+ H\phantom{^Z}\!\!\!
dx\wedge dt+E\cos[\omega (x-t)]dz\wedge dt\right)\\
&=-\left(Hdx+E\phantom{^Z}\!\!\! \cos[\omega(x-t)]dz
\right)\wedge(dt-dx)=-P\wedge Q. \end{align} If to $P$ we add $lQ$
($l$ being an arbitrary function) and use this sum $P'$ instead of
the former $P$, $\ast F$ does not change. It is obvious that
$P\cdot P<0$, but $P'\cdot P'=2lH-H^2-E^2\cos^2[\omega(t-x)] $.
Thus if we choose $l=H+\frac{E^2}{2H}\cos^2[\omega(t-x)]$, the
vector $P'$ will be timelike, $P'\cdot P'=H^2>0$, and we can take
$P'/H$ as a properly normalized monad, \be \label{Ptau} \tau=
\left(1+\frac{E^2 }{H^2}\cos^2 [\omega(t-x)]\right)(dt-dx)+
dx+\frac{E}{H} \cos[\omega(t-x)]dz. \ee Now the dually conjugated
field tensor reads $\ast F=-H\tau\wedge(dt-dx)$, thus in the frame
$\tau$ the electric field (\ref{elE}) vanishes, and this is the
field's co-moving frame. In all these calculations one has to
remember that when only one (here, magnetic) field survives after
the reference frame is transformed, there are other possible
transformations which do not change this situation (in fact, all
those which involve an additional motion in the direction of this
field, even when this motion occurs to be with a non-constant
magnitude of the three-velocity described by strictly local
Lorentz transformations working in non-inertial frames). Thus
there appears a continuum of such one-field frames ({\it cf.}
\cite{LanLif}, but working in general as well as in special
relativity), and the search for more elegant ones depends on the
individual taste of the researcher.

Let us now calculate the three-velocity of the frame $\tau$ from
the viewpoint of $\tau_{\mathrm{in}}$ using our general definition
(\ref{v}) and substitute the result into the left-hand side of
(\ref{emprop}), then putting into the right-hand side the
expressions of {\bf E} and {\bf B} from (\ref{sup}) in the frame
$\tau_{\mathrm{in}}$ to check if the Landau--Lifshitz definition
(\ref{emprop}) really works. Obviously, this way will not
represent a vicious circle since these parts of (\ref{emprop})
were initially deduced in \cite{LanLif} from a very different
standpoint than ours (moreover, in this way the left-hand side of
(\ref{Pauli}) will be automatically checked: both definitions of
{\bf v} cannot simultaneously work well). First, we rewrite
(\ref{v}) in these notations for frames and find
$\mathbf{v}\,(\perp\tau_{\mathrm{in}})$: \be \nonumber
\tau=(\tau\cdot \tau_{\mathrm{in}})(\tau_{\mathrm{in}}+\mathbf{v})
~ \Rightarrow ~ \mathbf{v}=\frac{\frac{E}{H}\cos[\omega(t-x)]}{1+
\frac{E^2}{H^2}\cos^2[\omega(t-x)]}\left(dz-\frac{E}{H}\cos
[\omega(t-x)]dx\right). \ee This means that \be \nonumber \frac{|
\mathbf{v}|}{1+\mathbf{v}^2}=\frac{\frac{E}{H}\cos[\omega(t-x)]
\sqrt{1+ \frac{E^2}{H^2}\cos^2[\omega(t-x)]}}{1+2\frac{E^2}{H^2}
\cos^2[\omega(t-x)]}. \ee Precisely the same is the result of
calculating $\frac{|\mathbf{E}\times\mathbf{B}|}{\mathbf{E}^2+
\mathbf{B}^2}$ --- Landau and Lifshitz's definition wins. (Pauli's
definition (\ref{Pauli}) cannot contain on its left-hand side the
construction $1+2\frac{E^2}{ H^2}\cos^2[\omega(t-x)]$ which
inevitably appears on the right-hand side
$2\frac{|\mathbf{E}\times\mathbf{B}|}{ \mathbf{E}^2+
\mathbf{B}^2}$, like in the Landau--Lifshitz case.) The mean value
of $|\mathbf{v}|$ is simply $\frac{2}{\pi}\arcsin\frac{E}{\sqrt{
E^2 + H^2}}$. When $H\rightarrow 0$, the mean propagation velocity
approaches that of light, while if $E\ll H$, the mean velocity can
become as low as one wishes: to this end, it is necessary to use
as strong magnetic field $H$ as possible and/or choose a
low-intensity wave in the superposition.

\section{Concluding remarks} \label{s12} \setcounter{equation}{0}

The results obtained in this paper are based on three simple
observations: that the physical classification of electromagnetic
fields should be formulated using the properties of only two well
known invariants of these fields, the complete description of
reference frame is related only to the state of motion of a
continuous multitude of test observers, and that the duality
rotation (in the vein of Rainich--Misner--Wheeler, but in a more
modern and general form) applied to a seed solution of Maxwell's
equations, yields a new solution in the same four-geometry which
was generated by the seed solution {\it via} Einstein's equations.
We have proven that these suppositions really work together, and
the duality rotation permits to {\it construct} qualitatively new
solutions, belonging also to other desired types of
electromagnetic fields in accordance with our classification.
There is only one restriction separating the pure null type fields
from those of other five types. The pure null type does not change
under the duality rotation, becoming in fact the same solution of
this pure type, though corresponding to another reference frame
and displaying the Doppler effect in its generalized form also
considered in this paper. As illustrations of application of our
approach we discuss concrete examples of the Kerr--Newman (KN)
solution and the Li\'enard--Wiechert (LW) field (to show the
efficiency of our method also in special relativity). Moreover, we
deduce three qualitatively new types of electromagnetic field
creating the same four-geometry as the seed KN solution, thus
describing other kinds of KN-like black holes. Studying the LW
field, we come upon a new conclusion that the linearity of
Maxwell's equations does not automatically mean that different
constituent parts of this field can be properly interpreted
separately. Other characteristics of the field (such as the energy
density and Poynting vector) have non-linear nature, thus a study
of these characteristics constructed only of one or another parts
of the LW solution, with omission of the combination
(``interaction'') of these parts, means a disregard of important
physical properties of the field, in particular, of its true
propagation velocity. We have explicitly shown that this velocity
of the complete LW solution is less than that of light, and we
have given the physically full-fledged frame co-moving with LW
field in which its Poynting vector exactly vanishes everywhere
outside the world line of the source of this field (strangely,
this fact was never noticed before). The last, but not least
example is related to a simple superposition of two exact
solutions of special-relativistic Maxwell's equations, plane
electromagnetic wave and homogeneous magnetic field in a vacuum.
We show that this super\-position, being itself an exact solution,
always propagates with the velocity lesser than that of light, and
we show that the elementary expression for this velocity is
properly defined in \cite{LanLif}, but not in \cite{Pauli}. (I
must admit that at first I liked the definition given in Pauli's
book much more than Landau--Lifshitz's one: see, {\it e.g.},
\cite{Mitsk06a}.)

Finally, may I express my hope (to a certain extent, against
hope), that the given here examples should lead our community of
physicists to a more profound consideration of the non-trivial
concept of reference frame and to its better understanding as a
more physical than purely mathematical subject and an important
ingredient in the description of physical reality. To console
those who cannot accept the representation of reference frames
through monads and Cartan's forms, I would add that they can take
instead any system of coordinates whose $t$-coordinate lines
coincide with those of the $\tau$-congruence (the choice of
spatial coordinates does not matter). In such a system, there will
be realized precisely the same picture, though mathematics will
feel awkward, while reference frames will seem to be silenced.

\end{document}